\documentclass[fleqn,12pt,twoside]{article}

\usepackage[headings]{espcrc1}
\usepackage{amsmath,amssymb}

\readRCS
$Id: espcrc1.tex,v 1.2 2004/02/24 11:22:11 spepping Exp $
\usepackage{graphicx}

\begin{document}

\title{
\begin{flushright}
\ \\*[-80pt] 
\begin{minipage}{0.2\linewidth}
\normalsize
KUNS-2126 \\*[50pt]
\end{minipage}
\end{flushright}
$D_4$ Flavor Symmetry for Neutrino Masses and Mixing}

\author{
Hajime Ishimori\address[1]{Graduate School of Science and Technology,
  Niigata University, Niigata 950-2181, Japan \\}%
    \thanks{e-mail: ishimori@muse.sc.niigata-u.ac.jp},
Tatsuo Kobayashi\address[2]{Department of Physics, Kyoto University, 
Kyoto 606-8502, Japan \\}
\thanks{e-mail: kobayash@gauge.scphys.kyoto-u.ac.jp},
Hiroshi Ohki\address[3]{Department of Physics, Kyoto University, 
Kyoto 606-8501, Japan \\}
\thanks{e-mail: ohki@scphys.kyoto-u.ac.jp}, 
Yuji Omura\addressmark[3]
\thanks{e-mail: omura@scphys.kyoto-u.ac.jp}, \\
Ryo Takahashi\addressmark[1]
\thanks{e-mail: takahasi@muse.sc.niigata-u.ac.jp},
Morimitsu Tanimoto\address[4]{Department of Physics, Niigata
  University, Niigata 950-2181, Japan}
\thanks{e-mail: tanimoto@muse.sc.niigata-u.ac.jp},\\}
       
\runtitle{$D_4$ flavor symmetry}
\runauthor{}

\maketitle

\begin{abstract}
 We  present the $D_4\times Z_2$  flavor symmetry,
which is different from the previous work  by Grimus and Lavoura.
Our model reduces to the standard model in the low energy and 
 there is no FCNC at the tree level.
Putting the experimental data,  parameters
are fixed, and then the implication  of our model is discussed.
 The condition to realize the tri-bimaximal mixing is presented.
The possibility for stringy realization of our model is also discussed.
\end{abstract}

\section{Introduction}

It is the important task to find  an origin of the observed hierarchies 
 in masses and flavor  mixing  for quarks and leptons.
Neutrino experimental  data provide us an important clue for this task. 
Especially, recent experiments of the neutrino
 oscillation go into the new  phase  of precise  determination of
 mixing angles and mass squared  differences \cite{Maltoni,Fogli}. 
Those indicate the  tri-bimaximal mixing  for three flavors 
 in the lepton sector \cite{HPS}. 
Therefore, it is necessary
to find a natural model that leads to this mixing pattern with good accuracy.

Flavor symmetries, in particular non-Abelian discrete flavor 
symmetries, are interesting ideas to realize realistic patterns of 
mass matrices.
Actually, several types of models with non-Abelian discrete flavor 
symmetries have been proposed \cite{Discrete}.
Furthermore, non-Abelian discrete flavor symmetries can be 
realized in the simple geometrical understanding of  
superstring theory \cite{Kobayashi:2004ya,Kobayashi:2006wq}
as well as extra dimensional models.
The $D_4$ symmetry can appear typically in heterotic string models 
on factorizable orbifolds including the $Z_2$ orbifold.
Indeed, several semi-realistic models with $D_4$ flavor symmetries have 
been constructed in Ref.~\cite{Kobayashi:2004ya,Kobayashi:2004ud} and 
in those models three families correspond to a singlet and 
a doublet under the $D_4$ flavor symmetry.
 Therefore, 
taking $D_4$ symmetry as the flavor symmetry of quarks and leptons,
 these  mass spectra and the flavor mixing matrix should be carefully
examined to establish the realistic model of  quarks and leptons
 \cite{Grimus,D4}.

 The  $D_4$ flavor symmetry was at first proposed 
for the  neutrino mass matrix  by Grimus and Lavoura \cite{Grimus}.
In this model, the atmospheric neutrino mixing is maximal while
 the solar neutrino mixing is arbitrary.
 They introduced three electroweak Higgs doublets together 
with two neutral singlets
 in the scalar sector to reproduce the large flavor mixing angles.
 Then, the tree level  flavor changing  neutral scalar vertices do not vanish.
Moreover,  when we consider supersymmetric extension of 
this $D_4$ flavor model, such a supersymmetric model would have 
three pairs of up and down Higgs fields.
That would violate the gauge coupling unification, 
which is one of important aspects of the minimal supersymmetric 
standard model, unless one introduces extra colored 
supermultiplets.\footnote{We would study a supersymmetric $D_4$ model 
in a separate paper \cite{SUSY-D4}.}

 In this paper, we propose alternative $D_4$ flavor model with 
 one  Higgs doublet, which reduces to the standard model in the low energy.
There is no tree level flavor changing neutral current (FCNC) in our model.
The higher dimensional operators provide the charged lepton 
and  neutrino masses.  Putting the experimental data, our parameters
are fixed, and then the implication  of our model is discussed.

The paper is organized as follows:
 we present the framework of the $D_4$ model in Sec. 2, and
 discuss the neutrino masses, flavor mixing angles  and Higgs potential 
in  Sec. 3. In Sec. 4, the numerical results are discussed.
Section 5 is devoted to the summary and discussion.

\section{$D_4$ flavor symmetry and Yukawa couplings}

 We present the framework of our $D_4$ flavor model.
The $D_4$ symmetry has five irreducible representations, 
that is, a doublet $2$ and four singlets, 
$1_{++}$, $1_{+-}$,  $1_{-+}$ and  $1_{--}$, where 
$1_{++}$ is a trivial singlet and the others are non-trivial singlets.
Their products are decomposed as 
\begin{equation}
2 \times 2 =  1_{++}+ 1_{+-}+ 1_{-+} + 1_{--}, \qquad 
1_{ab}\times 1_{cd} = 1_{ef},
\label{product}
\end{equation}
where $a,b,c,d= \pm$, $e=ac$ and $f=bd$.
Here,  the left-handed  lepton doublets  are denoted 
 as $l_\alpha \ (\alpha=e,\mu,\tau)$ and the right-handed charged leptons
and right-handed neutrinos are denoted as $e_R, \mu_R, \tau_R$, 
$N_e, N_\mu, N_\tau$, respectively.
The first family leptons are  assigned to $D_4$ trivial singlets, 
while second and third family ones are to $D_4$ doublets.
 The electroweak Higgs doublet $h$ is a $D_4$ trivial singlet.
We summarize charges of flavor symmetry in  Table \ref{charge}, where 
new  gauge singlet scalar  fields
$\chi$, $\chi_{-+}$, $\chi_1$, $\chi_2$ are introduced
 and  additional $Z_2$ charges
 are assigned for leptons and scalars.

\begin{table}[htb]
\begin{center}
\begin{tabular}{|c|cccccc||c||ccc|}
\hline
              &$l_e$   & $(l_\mu,l_\tau)$ & $e_R$  & $(\mu,\tau)_R$ & 
$N_e$ & $(N_\mu,N_\tau)$ &  $h$&$\chi$ &$\chi_{-+}$ &$(\chi_1,\chi_2)$   
\\ \hline
$D_4$      & $1_{++}$       & 2           & $1_{++}$  &     2 &  
     $1_{++}$ & 2    & $1_{++}$  & $1_{++}$ &$1_{-+}$&2   \\
$Z_2$ &    +    &   +     &  $-$ &      $-$     &   +     & +
& +  &$-$&$-$ & + \\ \hline
\end{tabular}
\end{center}
\caption{$D_4$ and $Z_2$ charges given for leptons and scalars.
\label{charge}}
\end{table}

\subsection{Charged lepton mass matrix}

We write down the Yukawa interactions, which are invariant under the gauge group of the standard model and the flavor symmetry $D_4\times Z_2$, by using the multiplication rule of $D_4$ in Eq. (\ref{product}),
\begin{eqnarray}
-{\cal L}_l
&=&y_e\bar el_e h\chi
+y_\tau(\bar\mu l_\mu+\bar\tau l_\tau)h\hat\chi
+y'_\tau(-\bar\mu l_\mu+\bar\tau l_\tau)h\hat\chi_{-+}
\nonumber\\&&
+y_{e\tau}\bar e( l_\mu\hat\chi_1+l_\tau\hat\chi_2)h\hat\chi
+y'_{e\tau}\bar e( l_\mu\hat\chi_1-l_\tau\hat\chi_2)h\hat\chi_{-+}
\nonumber\\&&
+y_{\tau e}(\bar\mu\hat\chi_1+\bar\tau\hat\chi_2)l_eh\hat\chi
+y'_{\tau e}(\bar\mu\hat\chi_1+\bar\tau\hat\chi_2)l_eh\hat\chi_{-+}
\nonumber\\&&
+y^a_{\mu\tau}(\bar\mu\hat\chi_1+\bar \tau\hat\chi_{2})(l_\mu\hat\chi_1+l_\tau\hat\chi_{2})h\hat\chi
+y^b_{\mu\tau}(\bar\mu\hat\chi_1-\bar \tau\hat\chi_{2})(l_\mu\hat\chi_1-l_\tau\hat\chi_{2})h\hat\chi
\nonumber\\&&
+y^c_{\mu\tau}(\bar\mu\hat\chi_2+ \bar \tau\hat\chi_{1})(l_\mu\hat\chi_2+ l_\tau\hat\chi_{1})h\hat\chi
+y^d_{\mu\tau}(\bar\mu\hat\chi_2- \bar \tau\hat\chi_{1})(l_\mu\hat\chi_2- l_\tau\hat\chi_{1})h\hat\chi
\nonumber\\&&
+y'{}^a_{\mu\tau}(\bar\mu\hat\chi_1+\bar \tau\hat\chi_{2})(l_\mu\hat\chi_1-l_\tau\hat\chi_{2})h\hat\chi_{-+}
+y'{}^b_{\mu\tau}(\bar\mu\hat\chi_1-\bar \tau\hat\chi_{2})(l_\mu\hat\chi_1+l_\tau\hat\chi_{2})h\hat\chi_{-+}
\nonumber\\&&
+y'{}^c_{\mu\tau}(\bar\mu\hat\chi_2+ \bar \tau\hat\chi_{1})(l_\mu\hat\chi_2- l_\tau\hat\chi_{1})h\hat\chi_{-+}
+y'{}^d_{\mu\tau}(\bar\mu\hat\chi_2- \bar \tau\hat\chi_{1})(l_\mu\hat\chi_2+ l_\tau\hat\chi_{1})h\hat\chi_{-+}
\nonumber\\&&
+\cdots+h.c.  ,
\label{lepton-yukawa-1}
\end{eqnarray}
where $\hat\chi$ and $\hat\chi_{-+}$ denote 
$\chi/\Lambda$ and $\chi_{-+}/\Lambda$ respectively, 
and $\Lambda$ is the cutoff scale. 
The  scale $\Lambda$ is taken to be the Planck one in our numerical study.
The ellipsis in Eq.~(\ref{lepton-yukawa-1}) denotes higher order contributions 
but they are negligibly small in our considerations.

We take the vacuum expectation values of scalar fields as follows: 
\begin{eqnarray}
\left< h\right>=v, \quad
\left< (\chi_1,\chi_2) \right>=(V_1,V_2), \quad
\left< \chi \right>=V_a, \quad
\left< \chi_{-+} \right>=V_b \ ,
\end{eqnarray}
where $v=174 {\rm GeV}$ and others are taken to be $D_4$ symmetry  
breaking scale.
After spontaneous symmetry breaking, 
the mass matrix of charged lepton becomes
\begin{eqnarray}
\label{chargedmass}
M_l
&=&v 
\left [
  \begin{array}{ccc}
y_e\alpha_a   &  (y_{e\mu}\alpha_a-y'_{e\mu}\alpha_b)\alpha & 
 (y_{e\mu}\alpha_a+y'_{e\mu}\alpha_b)\alpha\\ 
 (y_{\mu e}\alpha_a-y'_{\mu e}\alpha_b)\alpha  
  & y_\tau\alpha_a -y'_{\tau}\alpha_b  & 
 (y_{\mu\tau}\alpha_a+y'_{\mu\tau}\alpha_b)\alpha^2     \\
 (y_{\mu e}\alpha\alpha_a+y'_{\mu e}\alpha_b)\alpha    
&( y_{\mu\tau}\alpha_a-y'_{\mu\tau}\alpha_b)\alpha^2 &
 y_\tau\alpha_a+y'_{\tau}\alpha_b    \\  
\end{array} \right ],
\end{eqnarray}
where 
$\alpha_a\equiv V_a/\Lambda$ and 
$\alpha_b\equiv V_b/\Lambda$ and 
we assume the vacuum alignment in the $D_4$ doublet scalar field, 
$V_1=V_2$, so that, $\left< (\chi_1,\chi_2) \right>=(V,V)$. The 
parameter $\alpha$ is defined as $\alpha\equiv V/\Lambda$.  
This vacuum alignment is important for the masses and mixings in the neutrino sector. 
Since the value of $\alpha$ is
sufficiently small as discussed later, the charged lepton mass matrix can be 
approximately regarded as diagonal. 
The masses of 
charged leptons are given by 
\begin{eqnarray}
m_e=y_e\alpha_av,\quad
m_\mu=y_\tau\alpha_av-y'_\tau\alpha_bv,\quad
m_\tau=y_\tau\alpha_av+y'_\tau\alpha_bv.
\end{eqnarray}
We need the fine-tuning to obtain the difference between the masses of the muon and the tau, 
$m_\mu/m_\tau\ll1$, as discussed in Ref. \cite{Grimus}. 

\subsection{Neutrino mass matrix}

Let us consider the neutrino sector. We can write down the possible Dirac mass terms up to the dimension five 
operators by the same prescription as the charged lepton sector,

\begin{eqnarray}
-{\cal L}_D&=&
y_1\bar N_el_e \tilde h 
+y_2(\bar N_\mu l_\mu+\bar N_\tau l_\tau)\tilde h
\nonumber\\&&
+y_{12}\bar N_e( l_\mu\hat\chi_1+ l_\tau\hat\chi_2)\tilde h
+y_{21}(\bar N_\mu\hat\chi_1+N_\tau\hat\chi_2)l_e \tilde h
\nonumber\\&&
+y^a_{23}(\bar N_\mu\hat\chi_1+\bar  N_\tau\hat\chi_{2})(l_\mu\hat\chi_1+l_\tau\hat\chi_{2})\tilde h
\nonumber\\&&
+y^b_{23}(\bar N_\mu\hat\chi_1-\bar  N_\tau\hat\chi_{2})(l_\mu\hat\chi_1-l_\tau\hat\chi_{2})\tilde h
\nonumber\\&&
+y^c_{23}(\bar\mu\hat\chi_2+ \bar \tau\hat\chi_{1})(l_\mu\hat\chi_2+ l_\tau\hat\chi_{1})\tilde h
\nonumber\\&&
+y^d_{23}(\bar\mu\hat\chi_2- \bar \tau\hat\chi_{1})(l_\mu\hat\chi_2- l_\tau\hat\chi_{1})\tilde h
\nonumber\\&&
+\cdots+h.c. \ ,
\label{L-dirac}
\end{eqnarray}
where $\tilde h=i\tau_2 h^*$. 
The Majorana mass terms are given as 
\begin{eqnarray}
{\cal L}_R&=&
M_1N_e^TC^{-1}N_e
+M_2(N_\mu^TC^{-1}N_\mu+N_\tau^TC^{-1}N_\tau)
\nonumber\\&&
+y_aN_e^TC^{-1}( N_\mu\chi_1+N_\tau\chi_2)
\nonumber\\&&
+y^a_{b}(N_\mu^T\chi_1+N_\tau^T\chi_{2})C^{-1}( N_\mu\chi_1+N_\tau\chi_{2})/\Lambda
\nonumber\\&&
+y^b_{b}(N_\mu^T\chi_1-N_\tau^T\chi_{2})C^{-1}( N_\mu\chi_1-N_\tau\chi_{2})/\Lambda
\nonumber\\&&
+y^c_{b}(N_\mu^T\chi_2+ N^T_\tau\chi_{1})C^{-1}(N_\mu\chi_2+ N_\tau\chi_{1})/\Lambda
\nonumber\\&&
+y^d_{b}(N_\mu^T\chi_2- N_\tau^T\chi_{1})C^{-1}(N_\mu\chi_2- N_\tau\chi_{1})/\Lambda
\nonumber\\&&
+\cdots+h.c.
\label{L-majorana}
\end{eqnarray}

Then the neutrino mass matrices of Dirac $M_D$ and 
Majorana $M_R$ are given by
\begin{eqnarray}
M_D
&=&v 
\left(
  \begin{array}{ccc}
                  y_{1}    & y_{12}\alpha  & y_{12}\alpha  \\ 
                    y_{21}\alpha    & y_2  & y_{23}\alpha^2    \\
                    y_{21}\alpha     & y_{32}\alpha^2  & y_2   \\
  \end{array} \right),  \qquad
M_R
= 
\left(
  \begin{array}{ccc}
M_1   &  y_{a}\Lambda\alpha &  y_{a}\Lambda\alpha  \\ 
              y_{a}\Lambda\alpha    & M_2  & y_{b}\Lambda\alpha^2    \\
              y_{a}\Lambda\alpha   & y_{b}\Lambda\alpha^2 & M_2     \\  
\end{array} \right).
\end{eqnarray}
Similarly to the case of charged leptons, 
the ellipses in Eqs.~(\ref{L-dirac}) and (\ref{L-majorana}) correspond 
to higher order contributions 
but they are negligible.

The neutrino mass matrix is given by the see-saw mechanism,
\begin{eqnarray}
M_\nu=M_D M_R^{-1} M_D^T.
\end{eqnarray}
The neutrino mass matrix has the following structure,
\begin{eqnarray}
\label{massmatrix}
M_\nu
&\approx&v^2 
\left(
  \begin{array}{ccc}
A   &  B &  B \\ 
                     B    & C  & D     \\
                     B   & D & C    \\  
\end{array} \right),
\end{eqnarray}
where
\begin{eqnarray}
\label{AB}
A&=&
\frac{y_1^2M_2^2}{M_1M_2^2-2\alpha^2\Lambda^2M_2y_a^2},
\qquad
B=-
\frac{ y_1y_2 y_a\alpha\Lambda M_2}
{M_1M_2^2-2\alpha^2y_a^2\Lambda^2M_2},
\nonumber\\
C&=&
\frac{ y_2^2 (M_1M_2-y_a^2\alpha^2\Lambda^2)}
{M_1M_2^2-2\alpha^2y_a^2\Lambda^2M_2},
\qquad
D=
\frac{ y_2^2 y_a^2\alpha^2\Lambda^2}
{M_1M_2^2-2\alpha^2y_a^2\Lambda^2M_2}.
\end{eqnarray}
In these expressions, higher order terms are neglected
 under the assumption of 
\begin{equation}
 M_1 M_2^2 \gg \alpha^4\Lambda^3, \quad
M_2 \gg \alpha^2 M_1, \quad  M_2 \gg \alpha^2\Lambda, \quad \Lambda\gg M_1\ .
\label{appro}
\end{equation}
These assumptions are justified by the numerical analyses as discussed later.
The neutrino mass matrix is diagonalized by the following mixing matrix,
\begin{eqnarray}
V=\left(
  \begin{array}{ccc}
c   &  s &  0 \\ 
                     -s/\sqrt2    & c/\sqrt2  & 1/\sqrt2     \\
                     -s/\sqrt2   & c/\sqrt2 & -1/\sqrt2    \\  
\end{array} \right),
\end{eqnarray}
where $c\equiv \cos\theta_{12}$ and $s\equiv\sin\theta_{12}$ and $\theta_{12}$ corresponds to the solar mixing 
angle \cite{Grimus}. 
Then the neutrino mass matrix Eq.~(\ref{massmatrix}) 
is represented by the solar mixing and neutrino mass eigenvalues $m_i$ $(i=1\sim3)$ such as 
$M_\nu=V\mathrm{diag}(m_1,m_2,m_3)V^T$, which is 
\begin{eqnarray}
&&
\left(
  \begin{array}{ccc}
A   &  B &  B \\ 
                     B    & C  & D     \\
                      B   & D & C    \\  
\end{array} \right)v^2
\nonumber\\
&&
=   
\left(
  \begin{array}{ccc}
c^2m_1+s^2m_2   
&  -cs(m_1-m_2)/\sqrt2 
& -cs(m_1-m_2)/\sqrt2 
\\                      
-cs(m_1-m_2)/\sqrt2    
& (s^2m_1+c^2m_2+m_3)/2  
& (s^2m_1+c^2m_2-m_3)/2    
\\
-cs(m_1-m_2)/\sqrt2   
& (s^2m_1+c^2m_2-m_3)/2 
& (s^2m_1+c^2m_2+m_3)/2    \\  
\end{array} \right) \ ,
\nonumber\\
\end{eqnarray}
and we have the relations,
\begin{eqnarray}
&&Av^2=c^2m_1+s^2m_2,\quad
Bv^2=-\frac{cs}{\sqrt2}(m_1-m_2),\quad
\nonumber\\
&&Cv^2=\frac12(s^2m_1+c^2m_2+m_3),\quad
Dv^2=\frac12(s^2m_1+c^2m_2-m_3).
\end{eqnarray}
For neutrino masses, we find
\begin{eqnarray}
m_1+m_2&=&
\left(A+C+D\right)v^2 \ ,
\nonumber\\
m_1-m_2&=&
-\frac{\sqrt2}{cs}Bv^2 \ ,
\nonumber\\
m_3&=&
Cv^2-Dv^2\ .
\label{mass}
\end{eqnarray}
Then, 
the mass squared differences and the solar mixing angle are expressed by 
\begin{eqnarray}
\label{masstheta}
\Delta m_\mathrm{atm}^2
&=&-\frac14\left(A+C+D
-\frac{\sqrt2}{cs}B\right)^2v^4
+\left(C-D\right)^2v^4,
\nonumber\\
\Delta m_\mathrm{sol}^2
&=&\left(A+C+D\right)
\frac{\sqrt2}{cs}Bv^4,
\nonumber\\
\cot2\theta_{12}&=&
\frac{1}{2\sqrt2 B}
\left(C-A+D\right).
\end{eqnarray}

\subsection{Potential analysis}
Here, we analyze the scalar potential and discuss  the assumption of 
vacuum alignment, $V_1=V_2$. 
The relevant scalar potential
of ($\chi$, $\chi_{-+}$, $\chi_1$, $\chi_2$)
is given by
\begin{eqnarray}
-{\cal L}_v&=&
-\mu_1^2\chi^2-\mu_2^2\chi_{-+}^2-\mu_3^2(\chi_1^2+\chi_2^2)
\\&&
+\lambda_1\chi^4+\lambda_2\chi_{-+}^4
+\lambda_{3a}(\chi_1^2+\chi_2^2)^2+\lambda_{3b}(\chi_1^2-\chi_2^2)^2+
\lambda_{3c}\chi_1^2\chi_2^2
\nonumber\\&&
+\lambda_{12}\chi^2\chi_{-+}^2+\lambda_{13}\chi^2(\chi_1^2+\chi_2^2)+\lambda_{23}\chi_{-+}^2(\chi_1^2+\chi_2^2)
+\lambda_{123}\chi\chi_{-+}(\chi_1^2-\chi_2^2)\nonumber.
\end{eqnarray}
The minimum conditions are
\begin{eqnarray}
\frac{\partial{\cal L}_v}{\partial \chi}
&=&2V_a\left(
-\mu_1^2+2\lambda_1V_a^2+\lambda_{12}V_b^2
+\lambda_{13}(V_1^2+V_2^2)+\lambda_{123}\frac{V_b}{2V_a}(V_1^2-V_2^2)
\right)=0,
\nonumber\\
\frac{\partial{\cal L}_v}{\partial \chi_{-+}}
&=&2V_b\left(
-\mu_2^2+2\lambda_2V_b^2+\lambda_{12}V_a^2
+\lambda_{23}(V_1^2+V_2^2)+\lambda_{123}\frac{V_a}{2V_b}(V_1^2-V_2^2)
\right)
=0,
\nonumber\\
\frac{\partial{\cal L}_v}{\partial \chi_1}
&=&2V_1\left(\frac{}{}
-\mu_3^2+2\lambda_{3a}(V_1^2+V_2^2)
+2\lambda_{3b}(V_1^2-V_2^2)+\lambda_{3c}V_2^2
\right.
\nonumber\\&&
\left.
+\lambda_{13}V_a^2
+\lambda_{23}V_b^2+\lambda_{123}V_aV_b
\frac{}{}\right)
=0,
\nonumber\\
\frac{\partial{\cal L}_v}{\partial \chi_2}
&=&2V_2\left(\frac{}{}
-\mu_3^2+2\lambda_{3a}(V_1^2+V_2^2)
-2\lambda_{3b}(V_1^2-V_2^2)+\lambda_{3c}V_1^2
\right.
\nonumber\\&&
\left.
+\lambda_{13}V_a^2
+\lambda_{23}V_b^2-\lambda_{123}V_aV_b
\frac{}{}\right)
=0.
\end{eqnarray}
Since there are sixteen parameters ($\mu_{1,2,3}$, $\lambda_{1,2,3a,3b,3c}$, $\lambda_{12,13,23,123}$, 
$V_{a,b,1,2}$) while there are four equations, these equations 
can be solved. For this analysis, the following 
relation is important,
\begin{eqnarray}
(4\lambda_{3b}-\lambda_{3c})(V_1^2-V_2^2)
+2\lambda_{123}V_aV_b=0 ,
\end{eqnarray}
which is derived from $\partial{\cal L}_v/\partial \chi_1=0$ and 
$\partial{\cal L}_v/\partial \chi_2=0$.
To align the vacuum of $V_1=V_2$, 
one requires  $\lambda_{123}=0$, which is an assumption in our model.
We may impose additional symmetry to realize  $\lambda_{123}=0$.
 Inserting 
$\lambda_{123}=0$, we have 
\begin{eqnarray}
V_a^2&=& \frac{(2\lambda_2 \lambda_3-2\lambda_{23}^2)\mu_1^2 
+ (2\lambda_{13} \lambda_{23}-\lambda_{12}\lambda_{3})\mu_2^2 
+ (2\lambda_{12} \lambda_{23}-4\lambda_{2}\lambda_{13})\mu_3^2 }
{ 4\lambda_{1}\lambda_2\lambda_{3}
+4\lambda_{12}\lambda_{13}\lambda_{23} -
4\lambda_{1}\lambda_{23}^2 - 4\lambda_{2}\lambda_{13}^2 - 
\lambda_{3}\lambda_{12}^2 }, \nonumber \\
V_b^2&=& \frac{(2\lambda_{13} \lambda_{23}-\lambda_{12}\lambda_3)\mu_1^2 
+ (2\lambda_{1} \lambda_{3}-2\lambda_{13}^2)\mu_2^2 
+ (2\lambda_{12} \lambda_{13}-4\lambda_{1}\lambda_{23})\mu_3^2 }{
4\lambda_{1}\lambda_2\lambda_{3}
+4\lambda_{12}\lambda_{13}\lambda_{23} -
4\lambda_{1}\lambda_{23}^2 - 4\lambda_{2}\lambda_{13}^2 - 
\lambda_{3}\lambda_{12}^2}, \nonumber \\
V^2&=& \frac{(\lambda_{12} \lambda_{23}-2\lambda_2\lambda_{13})\mu_1^2 
+ (\lambda_{12} \lambda_{13}-2\lambda_{1}\lambda_{23})\mu_2^2 
+ (4\lambda_{1} \lambda_{2}-\lambda_{12}^2)\mu_3^2 }{
4\lambda_{1}\lambda_2\lambda_{3}
+4\lambda_{12}\lambda_{13}\lambda_{23} -
4\lambda_{1}\lambda_{23}^2 - 4\lambda_{2}\lambda_{13}^2 - 
\lambda_{3}\lambda_{12}^2}, 
\end{eqnarray}
where $\lambda_3 \equiv 4\lambda_{3a} + \lambda_{3c}$.
It is found that we can take $V_a\sim V_b$, which is necessary
to obtain muon and tau masses by adjusting   parameters. 

\section{Numerical discussion}
Let us discuss our numerical results. We define the following two dimensionless parameters, which are the ratios 
of $M_2$ and $\alpha\Lambda$ to $M_1$, respectively,
\begin{eqnarray}
 r\equiv \frac{M_2}{M_1} \ , \quad  k\equiv \frac{\alpha\Lambda}{M_1} \ .
\end{eqnarray} 
By using these parameters and Eq.~(\ref{AB}), the mass squared
differences 
and the solar mixing angle are rewritten
as
\begin{eqnarray}
\Delta m_\mathrm{atm}^2
&=&\frac{-(y_1^2r^2+y_2^2r+\sqrt2y_1y_2y_akr/cs)^2/4
+y_2^4(r-2y_a^2k^2)^2}
{(r^2-2y_a^2k^2r)^2}\cdot
\frac{v^4}{M_1^2} \ ,
\nonumber\\
\Delta m_\mathrm{sol}^2
&=&
\frac{-\sqrt2y_1y_2y_a k r (y_1^2r^2+y_2^2r)}
{(r^2-2y_a^2k^2r)^2cs}\cdot
\frac{v^4}{M_1^2} \ ,
\nonumber\\
\cot2\theta_{12}&=&
\frac{y_1^2r-y_2^2}
{2\sqrt2 y_1y_2y_ak}\ .
\label{mass-dif}
\end{eqnarray} 
The neutrino masses are given as 
\begin{eqnarray}
m_1&=&
\frac12\cdot
\frac{y_a^2r^2+y_2^2r+\sqrt2 y_1y_2y_akr/cs}
{r^2-2y_a^2k^2r}\times
\frac{v^2}{M_1} ,
\nonumber\\
m_2&=&
\frac12\cdot
\frac{y_a^2r^2+y_2^2r-\sqrt2 y_1y_2y_akr/cs}
{r^2-2y_a^2k^2r}\times
\frac{v^2}{M_1} ,
\nonumber\\
m_3&=&
\frac{y_2^2(r-2 y_a^2k^2)}
{r^2-2y_a^2k^2r}\times
\frac{v^2}{M_1}.
\label{mass2}
\end{eqnarray}

When we put the best fit values of mass squared differences and the solar mixing angle as 
$\Delta m_{\rm atm}^2=2.4\times10^{-3}$eV$^2$, $\Delta m_{\rm sol}^2=7.6\times10^{-5}$eV$^2$, and 
$\sin^2\theta_{12}=0.32$ \cite{Maltoni}, we have typical values of parameters in this model,
\begin{eqnarray}
 r =0.12, \qquad  k=-0.80, \qquad 
M_1= 4.9\times10^{15}{\rm GeV},
\label{value}
\end{eqnarray}
where we take all Yukawa couplings as $y_1=y_2=y_a=y_b=1$. 
By taking the cutoff scale 
  $\Lambda$ as the Planck scale $2.43 \times 10^{18}\ {\rm GeV}$, we find
\begin{eqnarray}
|\alpha|=1.6\times10^{-3}.
\end{eqnarray}
Therefore, the assumption to regard the diagonal mass 
matrix (\ref{chargedmass}) are  justified. 
The assumption of Eq.(\ref{appro}) turns to
\begin{eqnarray}
 |r| \gg |\alpha|^2, \quad  |r|^2\gg |\alpha k^3|, \quad |r|\gg |\alpha k|,
\quad  |k|\gg |\alpha| \ ,
\end{eqnarray}
which are also justified by the result in Eq.(\ref{value}).
The neutrino masses are given as
\begin{eqnarray}
m_1\sim
3.4{\rm meV},  \quad 
m_2\sim
-9.4{\rm meV}, \quad
m_3\sim
49{\rm meV} \ ,
\end{eqnarray}
which indicate the normal mass hierarchy.

\begin{figure}
\begin{center}
\includegraphics[scale = 1.0
]{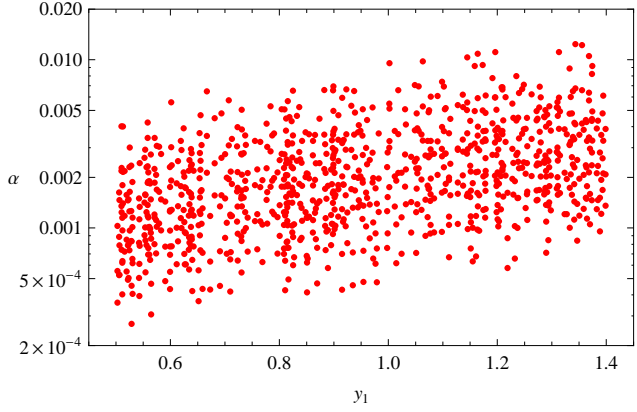}
\caption{Semilogarithmic plots for $\alpha$ versus $y_1$. 
\label{ran}}
\end{center}
\end{figure}

In the above numerical results, we have assumed 
all Yukawa couplings to be $1$.
Now let us consider how much the above results change 
by varying Yukawa couplings.
Following the above results, we assume that 
$|r| \ll 1$ and $ky_a = {\cal O}(1)$ for 
$y_1, y_2={\cal O}(1)$.\footnote{Note that either $k$ or $y_a$ 
can be small because only their product $ky_a$ appears 
in Eq.~(\ref{mass-dif}).}
Then, we approximate Eq.~(\ref{mass-dif}) as 
\begin{equation}
\Delta m_\mathrm{atm}^2
 \sim  \frac{y_2^4}{r^2} \frac{v^4}{M_1^2} \ , \qquad 
\Delta m_\mathrm{sol}^2
 \sim  - \frac{\sqrt{2}y_1y_2^3 }{4y_a^3k^3cs} \frac{v^4}{M_1^2} \ , 
\qquad 
\cot2\theta_{12}  \sim 
-\frac{y_2}{2\sqrt{2} y_1 y_a k}\ .
\label{mass-dif-2}
\end{equation} 
Hence, the parameters $k$, $r$, and $M_1$  are obtained as 
\begin{eqnarray}
 & & k \sim \frac{-1}{2\sqrt{2}}\frac{y_2}{y_1y_a}\tan 2\theta_{12} 
\sim 0.9 \times \frac{y_2}{y_1y_a} , 
\qquad r \sim \sqrt{\frac{\Delta m_\mathrm{sol}^2}{\Delta
    m_\mathrm{atm}^2}}
\left( \frac{y_2}{y_1}\right)^2 \sim 0.2 \times \left(
\frac{y_2}{y_1}\right)^2 , \nonumber \\
& & M_1 \sim 2\sqrt{2} v^2 \left( \frac{\cot^3 2\theta_{12}}{cs \Delta
    m_\mathrm{sol}^2} \right)^{1/2} y_1^2 
\sim 3\times 10^{15} \times y_1^2 {\rm~~GeV} , 
\end{eqnarray}
\noindent which leads to
\begin{eqnarray}
  \alpha = \frac{M_1 k}{\Lambda}\sim 0.001 \times \frac{y_1 y_2}{y_a} .
\end{eqnarray}
\noindent
The ratio $y_2/y_1$ must be of ${\cal O}(1)$ in order that 
the above approximation is valid, i.e. $y_ak={\cal O}(1)$.
Thus, values of $k$, $r$ and $M_1$ are of the same order as 
those in Eq.~(\ref{value}).
However, the value of $\alpha$ 
would change its order in some region even if we vary $y_1, y_2$ and $y_a$ 
by ${\cal O}(1)$, 
because $\alpha$ depends basically on a cube of ${\cal O}(1)$ 
parameters, i.e. $2^3 \sim 10$ and $0.5^3 \sim 0.1$.
Let us investigate this behavior numerically.
We use Eq.~(\ref{mass-dif}) and vary 
 $y_1$, $y_2$, and $y_a$ in the range of  $0.5-1.4$
 and taking account for the $3\sigma$ error-bar of input experimental data
$\Delta m^2_{\rm atm}$, $\Delta m^2_{\rm sol}$, and $\theta_{12}$.
We show the random plots of    $\alpha$ versus $y_1$ in Fig. \ref{ran}.
It is found  that  the value of  $\alpha$ is predicted 
around $10^{-4}- 10^{-2}$. 
The dependences of the value of $\alpha$ on other 
Yukawa couplings such as $y_2$ 
and $y_a$ are similar to the case of $y_1$. 
Thus we  obtain small $\alpha$ as long as 
Yukawa couplings are of ${\cal O}(1)$.

\section{Summary and Discussion}
We have presented the $D_4\times Z_2$  flavor symmetry,
which is different from the previous work  by Grimus and Lavoura.
Our model has  one Higgs doublet  although the neutrino mass matrix
 has the same structure as the one 
in the model by  Grimus and Lavoura. 
Our model reduces to the standard model in the low energy and 
 there is no FCNC at the tree level.

In order to realize the tri-bimaximal mixing,
 the condition $\cot 2\theta_{12}=\frac{1}{2\sqrt{2}}$ must be satisfied.
 Then, we have the condition $y_1^2 r-y_2^2=y_1y_2y_a k$.
Taking Yukawa couplings to be order one,
 this condition turns to simple one $r\simeq k+1$,
which is easily realized by adjusting parameters in our model.

It would be interesting to study supersymmetric extension 
of our model.
In such a supersymmetric $D_4$ model, we would have a 
specific pattern of superpartner mass matrices.
We would study it in a separate paper \cite{SUSY-D4}. 

Finally, we comment on the possibility for 
stringy realization of our model.
The $D_4$ flavor symmetry can be derived e.g. from 
heterotic string models on factorizable orbifolds including 
the $Z_2$ orbifold like $Z_2 \times Z_M$ 
orbifolds \cite{Kobayashi:2004ya,Kobayashi:2006wq}.
Indeed, several semi-realistic models have been 
constructed with three families
\cite{Kobayashi:2004ya,Kobayashi:2004ud}, 
where  three families consist of $D_4$ trivial singlets and doublets.
{}From this viewpoint, our $D_4$ flavor structure would  
be natural.
However, such orbifold models include only $D_4$ trivial singlets and 
doublets, but not $D_4$ non-trivial singlets as fundamental states.
The $D_4$ non-trivial singlet $\chi_{-+}$ 
plays an important role in our model.
We need to assume that $\chi_{-+}$ is a composite scalar of 
doublets, in order to obtain  $\chi_{-+}$ from the 
$Z_2$ orbifold.
Another possibility would be factorizable heterotic orbifold models 
including the $Z_4$ orbifold like $Z_4 \times Z_M$ orbifolds, 
because such orbifold models can lead to the $D_4$ flavor symmetry, 
where non-trivial singlets as well as trivial singlets 
and doublets can appear as fundamental modes \cite{Kobayashi:2006wq}.
Thus, it would be interesting to consider the realization 
of our $D_4$ model from $Z_4$ orbifold models.

\section*{Acknowledgments}

T.~K.\/ is supported in part by the
Grand-in-Aid for Scientific Research, No. 17540251 and 
the Grant-in-Aid for
the 21st Century COE ``The Center for Diversity and
Universality in Physics'' from the Ministry of Education, Culture,
Sports, Science and Technology of Japan.
The work of R.T. has been  supported by 
 Grand-in-Aid for  Scientific Research,  No.19$\cdot$4982 from
  the Japan Society of Promotion of Science.
The work of M.T. has been supported by the Grant-in-Aid for Science Research
from the Japan Society of Promotion of Science and
 the Ministry of Education, Science, and Culture of Japan,
Nos. 17540243 and 19034002.

\end{document}